\documentclass[conference]{IEEEtran}
\IEEEoverridecommandlockouts
\usepackage{cite}
\usepackage{amsmath,amssymb,amsfonts}
\usepackage{algorithmic}
\usepackage{graphicx}
\usepackage{textcomp}
\usepackage{xcolor}
\usepackage{tikz}
\usepackage{float}
\usepackage{url}
\usetikzlibrary{arrows.meta, positioning}
\def\BibTeX{{\rm B\kern-.05em{\sc i\kern-.025em b}\kern-.08em
    T\kern-.1667em\lower.7ex\hbox{E}\kern-.125emX}}
\begin{document}

\title{Data-driven Modeling of Grid-following Control in Grid-connected Converters 
\thanks{This material is based upon work supported by the National Science Foundation under Grant Number 2328241. Any opinions, findings, and conclusions or recommendations expressed in this material are those of the authors and do not necessarily reflect the views of the National Science Foundation.}
}


\author{
\IEEEauthorblockN{Amir~Bahador~Javadi}
\IEEEauthorblockA{\textit{Department of Electrical and Computer Engineering} \\
\textit{New Jersey Institute of Technology}\\
Newark, NJ, USA\\
\url{aj772@njit.edu}}
\and
\IEEEauthorblockN{Philip~Pong}
\IEEEauthorblockA{\textit{Department of Electrical and Computer Engineering} \\
\textit{New Jersey Institute of Technology}\\
Newark, NJ, USA\\
\url{philip.pong@njit.edu}}
}

\maketitle

\begin{abstract}


As power systems evolve with the integration of renewable energy sources and the implementation of smart grid technologies, there is an increasing need for flexible and scalable modeling approaches capable of accurately capturing the complex dynamics of modern grids. To meet this need, various methods, such as the sparse identification of nonlinear dynamics and deep symbolic regression, have been developed to identify dynamical systems directly from data. In this study, we examine the application of a converter-based resource as a replacement for a traditional generator within a lossless transmission line linked to an infinite bus system. This setup is used to generate synthetic data in grid-following control mode, enabling the evaluation of these methods in effectively capturing system dynamics.
\end{abstract}


\section{Introduction}

In recent years, the integration of renewable energy sources and power electronic converters into power grids has transformed the traditional dynamics and control paradigms. Power converters, especially grid-following and grid-forming converters \cite{khodam}, are central to this shift. These converters interface renewable energy sources with the grid but differ significantly in their control approaches and functionalities. Grid-following converters rely on the grid voltage for synchronization, using phase-locked loops to adjust their output to the grid frequency and phase, whereas grid-forming converters establish their own frequency and voltage reference, providing a source-like behavior for grid stability \cite{d2015virtual}.

Despite their essential role, grid-following converters pose several challenges, particularly under weak grid conditions or disturbances where phase-locked loops may induce instability. These stability issues arise due to the grid-following converter’s dependence on grid signals, making them vulnerable to frequency and phase disturbances. Additionally, the increasing penetration of grid-following converters, especially in low-inertia grids, has raised concerns about grid stability, power quality, and dynamic interactions, especially during faults and system transients \cite{huang2019grid}.

The modeling of grid-following converters is thus crucial for understanding their behavior under various grid conditions and for developing robust control strategies. Traditional physics-based models can be complex and may not capture all dynamics effectively, especially in rapidly changing grid conditions. Data-driven modeling approaches, like sparse identification of nonlinear dynamics (SINDy) and deep symbolic regression (DSR), offer promising alternatives by discovering governing equations directly from data \cite{JAVADI2025116075}. These methods can potentially reveal simplified, interpretable models that capture essential dynamics without the need for complex, predefined physical equations \cite{brunton2016discovering}.

SINDy, in particular, is effective for capturing sparsity in the governing dynamics by identifying the minimal set of terms that describe system behavior. It has been successfully applied to power system dynamics, making it a suitable candidate for grid-following converter modeling \cite{kaiser2018sparse}. DSR, leveraging advancements in machine learning and genetic programming, can further refine these models by exploring broader functional forms and nonlinearities, potentially improving predictive performance in dynamic environments \cite{petersen2019deep}. Together, these methods can contribute to improved modeling accuracy and control design for grid-following converters, facilitating their safe integration into modern power systems.

In this study, we applied both SINDy and DSR framework to a modified single machine infinite bus system, replacing the conventional generator with a converter-based resource. By simulating this system in grid-following control mode, we generate synthetic data to evaluate the efficacy of SINDy and DSR in capturing the system’s dynamics. Our results show that DSR provides a robust, interpretable model of the system, offering insights into the behavior of converter-based systems under this control strategy, even though this method takes more time to uncover the system dynamics compared to SINDy. This work contributes to the ongoing effort to develop advanced modeling tools capable of supporting the future grid’s stability and reliability.


\section{Grid-following Control of a stable Grid-connected Converter}
\label{sec2}
The bulk grid-connected system in this study is represented as a large, stable power grid, connected to the grid-following inverter through a lossless power grid transmission line. The resource is equipped with an LCL filter to reduce harmonic distortion and enhance the power quality of the output voltage. In this study, a single-machine infinite bus system is modified such that the conventional synchronous generator is replaced with a grid-following, converter-based resource, while maintaining the essential characteristics of the traditional system. The converter-based resource interfaces with an infinite bus, which is characterized by constant voltage and frequency \cite{pogaku2007modeling}.

\begin{figure}[h!]
\centering
\resizebox{0.45\textwidth}{!}{
\begin{tikzpicture}[
    block/.style = {draw, rectangle, minimum height=1cm, minimum width=2.5cm, align=center},
    line/.style = {draw, -},
    arrow/.style = {draw, -{Latex}},
    node distance = 1cm and 1cm
]

\node[block] (grid) {Bulk, and \\ stable grid};
\node[block, right=of grid] (line) {Transmission \\ line};
\node[block, right=of line] (lcl) {LCL filter};
\node[block, right=of lcl] (converter) {Converter};

\node[block, below=1cm of line] (outer) {Outer control};
\node[block, right=of outer] (inner) {Inner control};
\node[block, right=of inner] (pll) {PLL};

\draw[line] (grid) -- (line);
\draw[line] (line) -- (lcl);
\draw[line] (lcl) -- (converter);
\draw[line] (converter.south) -- ++(0,-0.8) -| (pll.north);

\draw[line] (outer) -- (inner);
\draw[line] (inner) -- (pll);

\draw[arrow] ([xshift=-2cm] outer.west) -- ++(2,0) node[midway, above] {$p^{ref}$, $q^{ref}$};
\end{tikzpicture}
}
\caption{Schematic of the grid-following control of the stable grid-connected converter.}
\label{CIB}
\end{figure}
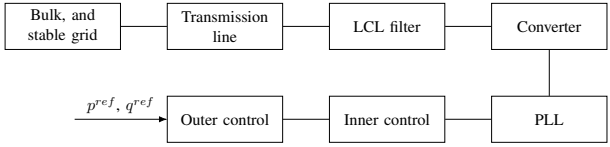

The diagram of the grid-following control of the stable grid-connected converter is presented in Fig. \ref{CIB}. In this diagram, the active ($p^{ref}$) and reactive power ($q^{ref}$) references are the control inputs. The voltages $\bar{v}_{2}$, $\bar{v}^{\textnormal{grid}}$, $\bar{v}^{\textnormal{filt}}$, and $\bar{v}^{\textnormal{cv}}$ correspond to the bulk and stable power grid, the transmission line, the LCL filter, and the converter, respectively. These are denoted as follows: $\bar{v}_{2} = v_{2}e^{j\theta_{2}}$, $\bar{v}^{\textnormal{grid}} = v_{r}^{\textnormal{grid}} + j v_{i}^{\textnormal{grid}}$, $\bar{v}^{\textnormal{filt}} = v_{r}^{\textnormal{filt}} + j v_{i}^{\textnormal{filt}}$, and $\bar{v}^{\textnormal{cv}} = v_{r}^{\textnormal{cv}} + j v_{i}^{\textnormal{cv}}$.
The inductance leg of the LCL filter on the transmission side is represented by $r_{g} + j\omega\ell_{g}$, while the inductance leg on the converter side is represented by $r_{f} + j\omega\ell_{f}$. Additionally, $c_{f}$ denotes the capacitance of the LCL filter.
The power flow between the grid-following converter, the bulk, and the stable grid, as well as the dynamics of the LCL filter and transmission line, are governed by~\eqref{1} and \eqref{2}. Furthermore, the dynamics of the converter are modeled using the average model, as shown in~\eqref{eq3}.

Grid-following control mode is a critical strategy for converters operating in interconnected power systems, enabling them to synchronize with the grid and maintain stability amidst varying operational conditions. This mode leverages a phase-locked loop (PLL) to precisely track the grid's voltage phase and frequency, ensuring that the converter's output is well-aligned with the grid's parameters. The PLL dynamically adjusts the converter's internal voltage references based on filtered voltage signals, allowing it to respond rapidly to changes in grid conditions. Complementing this, outer control loop regulates active and reactive power by continuously comparing reference power levels with actual measurements, effectively managing power flow and enhancing grid stability. Inner control loop further refines performance by controlling current output to minimize deviations, ensuring that the converter operates efficiently and responsively. Through this multi-layered control approach, grid-following converters play a vital role in the reliable integration of renewable energy sources into the power grid, contributing to a more resilient and flexible energy landscape \cite{aljarrah2024issues}.

\begin{align}
\textnormal{Power flow}
&\begin{cases}
\bar{v}_{2} &=v_{2}e^{j\theta_{2}} \\
\bar{v}^{\textnormal{grid}} &=v^{\textnormal{grid}}e^{j\theta^{\textnormal{grid}}} = v_{r}^{\textnormal{grid}} + j v_{i}^{\textnormal{grid}} \\
\bar{v}^{\textnormal{filt}} &=v^{\textnormal{filt}}e^{j\theta^{\textnormal{filt}}} = v_{r}^{\textnormal{filt}} + j v_{i}^{\textnormal{filt}} \\
0&= \left(v^{\textnormal{grid}}\right)^{2}G_{\textnormal{11}} + v^{\textnormal{grid}}v_{2}G_{\textnormal{12}}\cos{\left(\theta^{\textnormal{grid}}-\theta_{2}\right)} \\&+ v^{\textnormal{grid}}v_{2}B_{\textnormal{12}}\sin{\left(\theta^{\textnormal{grid}}-\theta_{2}\right)} +\left(v^{\textnormal{grid}}\right)^{2}G_{\textnormal{ff}} \\&+ v^{\textnormal{grid}}v^{\textnormal{filt}}G_{\textnormal{1f}}\cos{\left(\theta^{\textnormal{grid}}-\theta^{\textnormal{filt}}\right)} \\&+ v^{\textnormal{grid}}v^{\textnormal{filt}}B_{\textnormal{1f}}\sin{\left(\theta^{\textnormal{grid}}-\theta^{\textnormal{filt}}\right)}\\
Y_{\textnormal{1}} &= G_{\textnormal{11}} + jB_{\textnormal{11}} = \frac{1}{R+jX} \\
G_{\textnormal{12}} &= -G_{\textnormal{11}}, \\
B_{\textnormal{12}} &= -B_{\textnormal{11}} \\
Y_{\textnormal{f}} &= G_{\textnormal{ff}} + jB_{\textnormal{ff}} = \frac{1}{r_{g}+j\omega\ell_{g}} \\
G_{\textnormal{1f}} &= -G_{\textnormal{ff}}, \\
B_{\textnormal{1f}} &= -B_{\textnormal{ff}}
\label{1}
\end{cases}\\
\textnormal{LCL filter}
&\begin{cases}
\frac{\ell_{f}}{\omega_{b}} \frac{d}{dt}{i}_{r}^{\textnormal{cv}}&= v_{r}^{\textnormal{cv}} - v_{r}^{\textnormal{filt}} - r_{f}i_{r}^{\textnormal{cv}} + \omega_{s}\ell_{f}i_{i}^{\textnormal{cv}} \\
\frac{\ell_{f}}{\omega_{b}} \frac{d}{dt}{i}_{i}^{\textnormal{cv}}&= v_{i}^{\textnormal{cv}} - v_{i}^{\textnormal{filt}} - r_{f}i_{i}^{\textnormal{cv}} - \omega_{s}\ell_{f}i_{r}^{\textnormal{cv}} \\
\frac{c_{f}}{\omega_{b}} \frac{d}{dt}{v}_{r}^{\textnormal{filt}}&= i_{r}^{\textnormal{cv}} - i_{r}^{\textnormal{filt}} + \omega_{s}c_{f}v_{i}^{\textnormal{filt}} \\
\frac{c_{f}}{\omega_{b}} \frac{d}{dt}{v}_{i}^{\textnormal{filt}}&= i_{i}^{\textnormal{cv}} - i_{i}^{\textnormal{filt}} - \omega_{s}c_{f}v_{r}^{\textnormal{filt}} \\
\frac{\ell_{g}}{\omega_{b}} \frac{d}{dt}{i}_{r}^{\textnormal{filt}}&= v_{r}^{\textnormal{filt}} - v_{r}^{\textnormal{grid}} - r_{g}i_{r}^{\textnormal{filt}} + \omega_{s}\ell_{g}i_{i}^{\textnormal{filt}} \\
\frac{\ell_{g}}{\omega_{b}} \frac{d}{dt}{i}_{i}^{\textnormal{filt}}&= v_{i}^{\textnormal{filt}} - v_{i}^{\textnormal{grid}} - r_{g}i_{i}^{\textnormal{filt}} - \omega_{s}\ell_{g}i_{r}^{\textnormal{filt}} \\
v_{d}^{\textnormal{filt}}&= \cos{(\theta^{\textnormal{pll}}})v_{r}^{\textnormal{filt}} + \sin{(\theta^{\textnormal{pll}}})v_{i}^{\textnormal{filt}} \\
v_{q}^{\textnormal{filt}}&= -\sin{(\theta^{\textnormal{pll}}})v_{r}^{\textnormal{filt}} + \cos{(\theta^{\textnormal{pll}}})v_{i}^{\textnormal{filt}}\\
%
%
%
i_{d}^{\textnormal{cv}}&= \cos{(\theta^{\textnormal{pll}}})i_{r}^{\textnormal{cv}} + \sin{(\theta^{\textnormal{pll}}})i_{i}^{\textnormal{cv}} \\
i_{q}^{\textnormal{cv}}&= -\sin{(\theta^{\textnormal{pll}}})i_{r}^{\textnormal{cv}} + \cos{(\theta^{\textnormal{pll}}})i_{i}^{\textnormal{cv}}
\label{2}
\end{cases}\\
\textnormal{Converter}
&\begin{cases}
\bar{v}^{\textnormal{cv}} &=v^{\textnormal{cv}}e^{j\theta^{\textnormal{cv}}} = v_{r}^{\textnormal{cv}} + j v_{i}^{\textnormal{cv}}\\
v_{r}^{\textnormal{cv}} &= \cos{(\theta^{\textnormal{pll}}})v_{d}^{\textnormal{cv,ref}} - \sin{(\theta^{\textnormal{pll}}})v_{q}^{\textnormal{cv,ref}} 
\\
v_{i}^{\textnormal{cv}} &= \sin{(\theta^{\textnormal{pll}}})v_{d}^{\textnormal{cv,ref}} + \cos{(\theta^{\textnormal{pll}}})v_{q}^{\textnormal{cv,ref}}
\label{eq3}
\end{cases}
\end{align}


The dynamics of the system are governed by a series of interconnected equations that ensure the converter synchronizes with the grid and maintains stability. The dynamics associated with the PLL, outer and inner control loops are represented in~\eqref{5}, \eqref{6}, and \eqref{7}, respectively. These sets of equations are consistent with the LCL filter, power grid, and the average model of the converter, which have been utilized to generate data for evaluating the performance of both SINDy and DSR frameworks, with the goal of directly uncovering the underlying dynamics from data.

\begin{align}
\textnormal{PLL}
&\begin{cases}
\frac{1}{\ell_{lp}} \frac{d}{dt}{v}_{q}^{\textnormal{pll}} &= v_{q}^{\textnormal{filt}} - v_{q}^{\textnormal{pll}} \\
\frac{d}{dt}{\epsilon}^{\textnormal{pll}} &= v_{q}^{\textnormal{pll}} \\
\frac{1}{\omega_{b}} \frac{d}{dt}{\theta}^{\textnormal{pll}} &= k_{p}^{\textnormal{pll}}v_{q}^{\textnormal{pll}} + k_{i}^{\textnormal{pll}}\epsilon^{\textnormal{pll}} + 1 - \omega_s \\
\omega^{\textnormal{pll}} &= k_{p}^{\textnormal{pll}}v_{q}^{\textnormal{pll}} + k_{i}^{\textnormal{pll}}\epsilon^{\textnormal{pll}} + 1
\label{5}
\end{cases}
\end{align}

\begin{align}
\textnormal{Outer control}
&\begin{cases}
\frac{d}{dt}{\sigma}_{p} &= p^{\textnormal{ref}} - p_{m} \\
\frac{1}{\omega_{z}} \frac{d}{dt}{p}_{m} &= v_{r}^{\textnormal{filt}}i_{r}^{\textnormal{filt}} + v_{i}^{\textnormal{filt}}i_{i}^{\textnormal{filt}} - p_{m} \\
\frac{d}{dt}{\sigma}_{q} &= q^{\textnormal{ref}} - q_{m} \\
\frac{1}{\omega_{f}} \frac{d}{dt}{q}_{m} &= -v_{r}^{\textnormal{filt}}i_{i}^{\textnormal{filt}} + v_{i}^{\textnormal{filt}}i_{r}^{\textnormal{filt}} - q_{m} \\
i_{q}^{\textnormal{ref}} &= k_{p}^{p}\left(p^{\textnormal{ref}} - p_{m}\right) + k_{i}^{p}\sigma_{p} \\
i_{d}^{\textnormal{ref}} &= k_{p}^{q}\left(q^{\textnormal{ref}} - q_{m}\right) + k_{i}^{q}\sigma_{q}
\label{6}
\end{cases}\\
\textnormal{Inner control}
&\begin{cases}
\frac{d}{dt}{\gamma}_{d} &= i_{d}^{\textnormal{ref}} - i_{d}^{\textnormal{cv}} \\
\frac{d}{dt}{\gamma}_{q} &= i_{q}^{\textnormal{ref}} - i_{q}^{\textnormal{cv}} \\
v_{d}^{\textnormal{ref}} &= k_{p}^{c} \left(i_{d}^{\textnormal{ref}} - i_{d}^{\textnormal{cv}}\right) + k_{i}^{c}\gamma_{d} \\
v_{q}^{\textnormal{ref}} &= k_{p}^{c} \left(i_{q}^{\textnormal{ref}} - i_{q}^{\textnormal{cv}}\right) + k_{i}^{c}\gamma_{q} \\
v_{d}^{\textnormal{cv,ref}} &= v_{d}^{\textnormal{ref}} - \omega^{\textnormal{pll}}\ell_{f}i_{q}^{\textnormal{cv}} + k_{\textnormal{ffv}}v_{d}^{\textnormal{filt}} \\
v_{q}^{\textnormal{cv,ref}} &= v_{q}^{\textnormal{ref}} + \omega^{\textnormal{pll}}\ell_{f}i_{d}^{\textnormal{cv}} + k_{\textnormal{ffv}}v_{q}^{\textnormal{filt}}
\label{7}
\end{cases}
\end{align}

\section{Sparse Identification of Nonlinear Dynamics (SINDy)}

In the context of power systems, SINDy can be used to identify the dynamics of grid-forming \cite{khodam} or grid-following inverter systems. These systems are critical for modern electrical grids, where inverters are used to convert DC to AC and manage the flow of power. By understanding the dynamics of the inverters, SINDy can inform control strategies for voltage regulation, frequency stabilization, and disturbance rejection \cite{brunton2016discovering}.

SINDy is based on a regression approach where the system dynamics are assumed to be represented by a sparse combination of basis functions \cite{kaiser2018sparse}. The general form of the dynamical system can be written as:
\[
\frac{d\mathbf{x}}{dt} = f(\mathbf{x}, \mathbf{u}, t) = \sum_{i=1}^{N} \theta_i \varphi_i(\mathbf{x}, \mathbf{u}, t),
\]
where $\mathbf{x}$ represent the system states, $\mathbf{u}$ are the control inputs, $\varphi_i$ are a set of candidate basis functions, and $\theta_i$ are the coefficients to be determined. The objective of SINDy is to determine the sparse vector $\theta = [\theta_1, \theta_2, \dots, \theta_N]$ that best approximates the system's behavior from observed data.

Incorporating SINDy with control systems allows for the derivation of control laws based on the identified system dynamics. One common approach is to use the identified dynamics in model predictive control or optimal control frameworks. By estimating the system's evolution, control inputs can be adjusted to maintain system stability and optimize performance \cite{zhang2019convergence}.

\section{Deep Symbolic Regression (DSR)}

In power systems, DSR can be applied to model the dynamics of inverter-based systems such as grid-following and grid-forming inverters \cite{khodam}. These systems have complex dynamics due to nonlinearities, time-varying parameters, and external disturbances. The identified models from DSR can inform control strategies for voltage regulation, frequency stabilization, and disturbance rejection, which are essential for the reliable operation of modern power grids. For example, in grid-following inverters, DSR can be used to model the inverter's response to changes in reference power or voltage. These models can then be used in controller design to maintain system stability and improve the efficiency of power flow management. Similarly, for grid-forming inverters, which are designed to provide grid support, DSR-based control designs can help stabilize grid voltage and frequency \cite{petersen2019deep}.

DSR combines deep learning with symbolic regression to automatically discover governing equations from data. It involves training a neural network to propose candidate equations for a given dynamical system. The network learns to generate symbolic expressions based on data input, and the most promising models are then selected and refined through evolutionary algorithms or optimization techniques \cite{biggio2021neural}. The general form of a discovered model might look like:
\[
\frac{d\mathbf{x}}{dt} = f(\mathbf{x}, \mathbf{u} , t) = \sum_{i=1}^{N} g_i(\mathbf{x}, \mathbf{u}, t),
\]

where $\mathbf{x}$ represents the system state, $\mathbf{u}$ are the control inputs, and $g_i$ represent symbolic expressions learned from the data. DSR allows for the discovery of more complex, nonlinear relationships between system states, which is especially useful in control systems where understanding the exact form of system dynamics is critical for controller design. The discovered symbolic models can be directly used in model-based control strategies such as model predictive control or optimal control. By using the learned dynamics of the system, control inputs can be optimized to achieve desired system behavior while accounting for disturbances or parameter variations \cite{khodam}.

\section{Numerical Results \& Discussion}

In this study, we utilized the dynamics presented in section \ref{sec2} which discusses the grid-following control mode, to generate a synthetic dataset in MATLAB. It is noteworthy that during the data collection phase, two different disturbances were introduced to the system. The first event
occurred at \(t=0.3\)s, involving a change in \(p^{ref}\) to 0.7 p.u., and the second event
occurred at \(t=0.6\)s, involving a change in \(q^{ref}\) to 0.2 p.u. This dataset was used to assess the effectiveness of both the SINDy \cite{kaiser2018sparse} and DSR \cite{petersen2019deep} methods in accurately capturing the dynamics associated with this control mode. Following the data generation process, we implemented both the SINDy \cite{brunton2016sparse, kaptanoglu2021pysindy} and DSR \cite{landajuela2022unified} methodologies for model identification of the system. 
The specific configurations and parameters of the system under study, expressed in per unit, are as follows: $X$=0.0020625, \(\ell_f\)=0.009, $r_f$=0.016, $c_f$=2.5, $\ell_g$=0.002, and $r_g$=0.003. 

\subsection{LCL filter}

\begin{figure*}[htbp]
    \centering
    \includegraphics[width=1\linewidth]{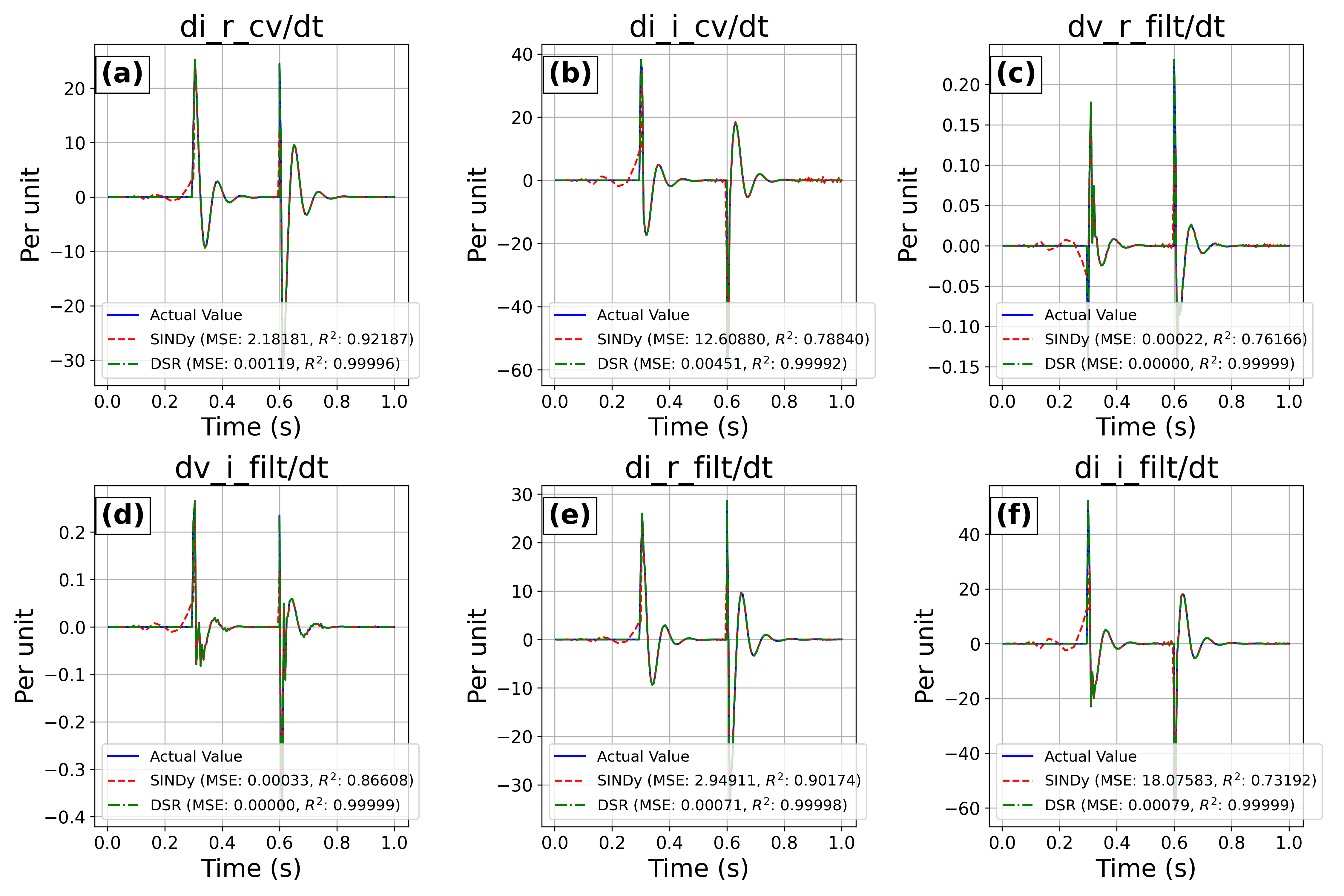}
    \caption{Comparison of the identified dynamics of the LCL filter for the grid-connected converter in grid-following control mode, using the SINDy and DSR methods. Each subplot illustrates the time derivative of state variables under two distinct disturbances: a change in active power reference (\(p^{ref} = 0.7\)) at \(t=0.3\)s and a change in reactive power reference (\(q^{ref} = 0.2\)) at \(t=0.6\)s. The subplots correspond to each part are as follow; (a)$\frac{d}{dt}{i}_{r}^{\textnormal{cv}}$, (b)$\frac{d}{dt}{i}_{i}^{\textnormal{cv}}$, (c)$\frac{d}{dt}{v}_{r}^{\textnormal{filt}}$, (d)$\frac{d}{dt}{v}_{i}^{\textnormal{filt}}$, (e)$\frac{d}{dt}{i}_{r}^{\textnormal{filt}}$, and (f)$\frac{d}{dt}{i}_{i}^{\textnormal{filt}}$.}
    \label{filter}
\end{figure*}

The performance of both methods corresponding to each state related to the LCL filter is illustrated in Figs. \ref{filter}.a, \ref{filter}.b, \ref{filter}.c, \ref{filter}.d, \ref{filter}.e, and \ref{filter}.f. SINDy model captures the general trend of the system’s response for the first state following the disturbances at \( t = 0.3 \, \text{s} \) and \( t = 0.6 \, \text{s} \) in Fig. \ref{filter}.a, but exhibits a relatively high mean square error and an \( R^2 \) score of 0.92, indicating some deviation from the actual values. Conversely, the DSR model closely follows the actual values, achieving a notably low mean square error of 0.001 and a high \( R^2 \) score of 0.99, demonstrating its accuracy in modeling this state. A significant discrepancy between the SINDy model and the actual values is observed in Fig. \ref{filter}.b, as evidenced by the high mean square error of 12.6 and a low \( R^2 \) score of 0.78. The DSR model, however, performs remarkably well, with a mean square error of 0.004 and an \( R^2 \) score of 0.99, indicating its robustness in handling this system response. 

The SINDy model shows improved accuracy in Fig. \ref{filter}.c, compared to Fig. \ref{filter}.b, with a mean square error of 0.0002 and an \( R^2 \) score of 0.76. Despite this, DSR provides a more accurate model, achieving a low mean square error and an \( R^2 \) score of 0.99, further validating its capability in capturing the dynamics accurately. The performance of the SINDy model in Fig. \ref{filter}.d is characterized by a mean square error of 0.0003 and an \( R^2 \) score of 0.86, indicating reasonable alignment with the actual values. However, DSR still achieves a lower mean square error and a higher \( R^2 \) score of 0.99, suggesting a more accurate model. 

The SINDy model achieves a mean square error of 2.94 and an \( R^2 \) score of 0.9 in Fig. \ref{filter}.e, reflecting low accuracy. However, the DSR model continues to exhibit a more precise fit, with a mean square error of 0.0007 and an \( R^2 \) score of 0.99, reinforcing its effectiveness in this scenario. Finally, the SINDy model displays a relatively high mean square error of 18.07 and a low \( R^2 \) score of 0.73 in Fig. \ref{filter}.f, indicating significant deviation from the actual values. The DSR model again demonstrates more accurate performance, with a mean square error of 0.00079 and an \( R^2 \) score of 0.99.

\begin{figure*}[ht]
    \centering
    \includegraphics[width=1\linewidth]{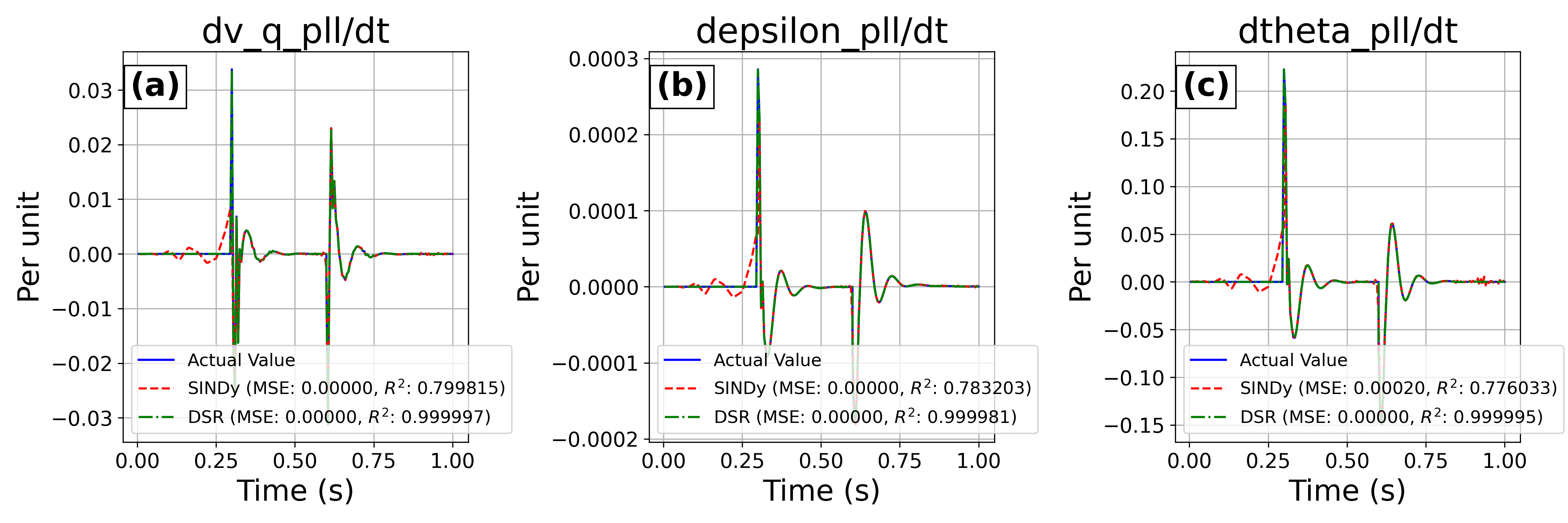}
    \caption{Comparison of the identified dynamics of the PLL in the grid-connected converter under grid-following control mode, using the SINDy and DSR methods. Each subplot illustrates the time derivative of state variables under two distinct disturbances: a change in active power reference (\(p^{ref} = 0.7\)) at \(t=0.3\)s and a change in reactive power reference (\(q^{ref} = 0.2\)) at \(t=0.6\)s. The subplots correspond to each part are as follow; (a)$\frac{d}{dt}{v}_{q}^{\textnormal{pll}}$,  (b)$\frac{d}{dt}{\epsilon}^{\textnormal{pll}}$, and (c)$\frac{d}{dt}{\theta}^{\textnormal{pll}}$}
    \label{pll}
\end{figure*}

\subsection{PLL}
The discovered dynamics related to the PLL states are presented in Figs. \ref{pll}.a, \ref{pll}.b, and \ref{pll}.c. The SINDy model attains an \( R^2 \) score of 0.79 in Fig. \ref{pll}.a. While SINDy demonstrates moderate performance, DSR achieves an \( R^2 \) score of 0.99, indicating a highly accurate fit to the actual values. Similarly, the SINDy model shows an \( R^2 \) score of 0.78 in Fig. \ref{pll}.b, demonstrating moderate accuracy. In contrast, DSR achieves an \( R^2 \) score close to 1, with a low mean square error, signifying its precision. The performance of the SINDy model with a mean square error of 0.0002 and an \( R^2 \) score of 0.77 is illustrated in Fig. \ref{pll}.c, indicating fair alignment with the actual values. Nevertheless, DSR once again captures the dynamics more accurately, with an mean square error close to zero and an \( R^2 \) score of 0.99.

\subsection{Outer control}
\begin{figure*}
    \centering
    \includegraphics[width=1\linewidth]{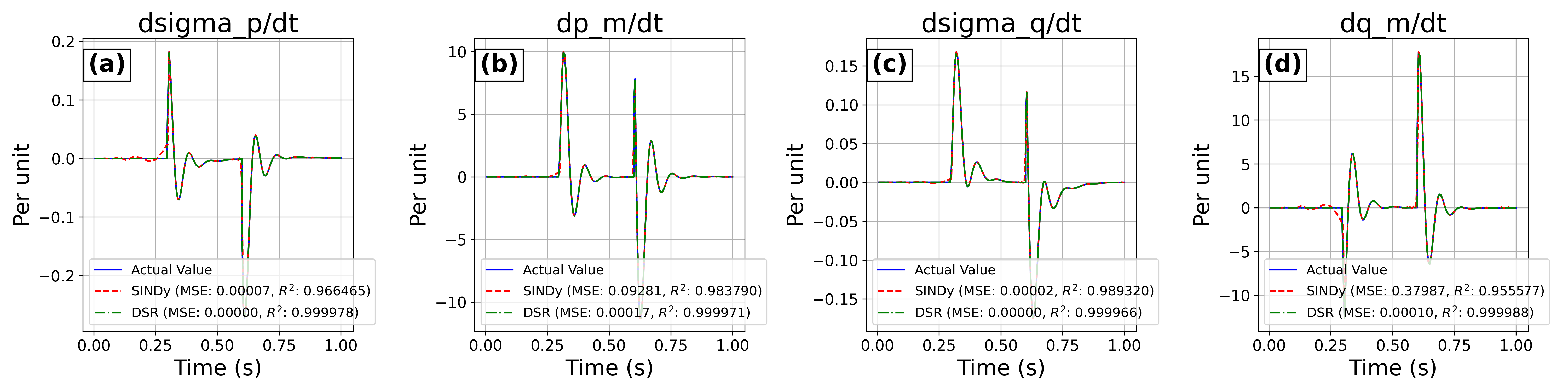}
    \caption{Comparison of the identified dynamics of the outer control in the grid-connected converter under grid-following control mode, using the SINDy and DSR methods. Each subplot illustrates the time derivative of state variables under two distinct disturbances: a change in active power reference (\(p^{ref} = 0.7\)) at \(t=0.3\)s and a change in reactive power reference (\(q^{ref} = 0.2\)) at \(t=0.6\)s. The subplots correspond to each part are as follow; (a)$\frac{d}{dt}{\sigma}_{p}$,  (b)$\frac{d}{dt}{p}_{m}$, (c)$\frac{d}{dt}{\sigma}_{q}$, and (d)$\frac{d}{dt}{q}_{m}$}
    \label{outer}
\end{figure*}

The identified model from both methodologies for outer control are indicated in Figs. \ref{outer}.a, \ref{outer}.b, \ref{outer}.c, and \ref{outer}.d. SINDy model achieves a mean square error of 0.00007 and an \( R^2 \) score of 0.96 in Fig. \ref{outer}.a, indicating a reasonable fit. In contrast, DSR demonstrates near-perfect alignment, with an mean square error close to zero and an \( R^2 \) score of 0.99. The SINDy model in Fig. \ref{outer}.b has a mean square error of 0.09 and an \( R^2 \) score of 0.98, reflecting strong performance. Although SINDy performs well, DSR achieves a more accurate model with a mean square error of 0.0001 and an \( R^2 \) score of 0.99. The SINDy model exhibits a mean square error of 0.00002 and an \( R^2 \) score of 0.98 in Fig. \ref{outer}.c, demonstrating high accuracy. Nevertheless, DSR continues to capture the system behavior more precisely, with a mean square error close to zero and an \( R^2 \) score of 0.99. Finally, SINDy attains a mean square error of 0.37 and an \( R^2 \) score of 0.95 in Fig. \ref{outer}.d, indicating satisfactory performance. DSR, however, provides a closer fit to the actual values with a mean square error of 0.0001 and an \( R^2 \) score of 0.99. 

\begin{figure}[!hb]
    \centering
    \includegraphics[width=1\linewidth]{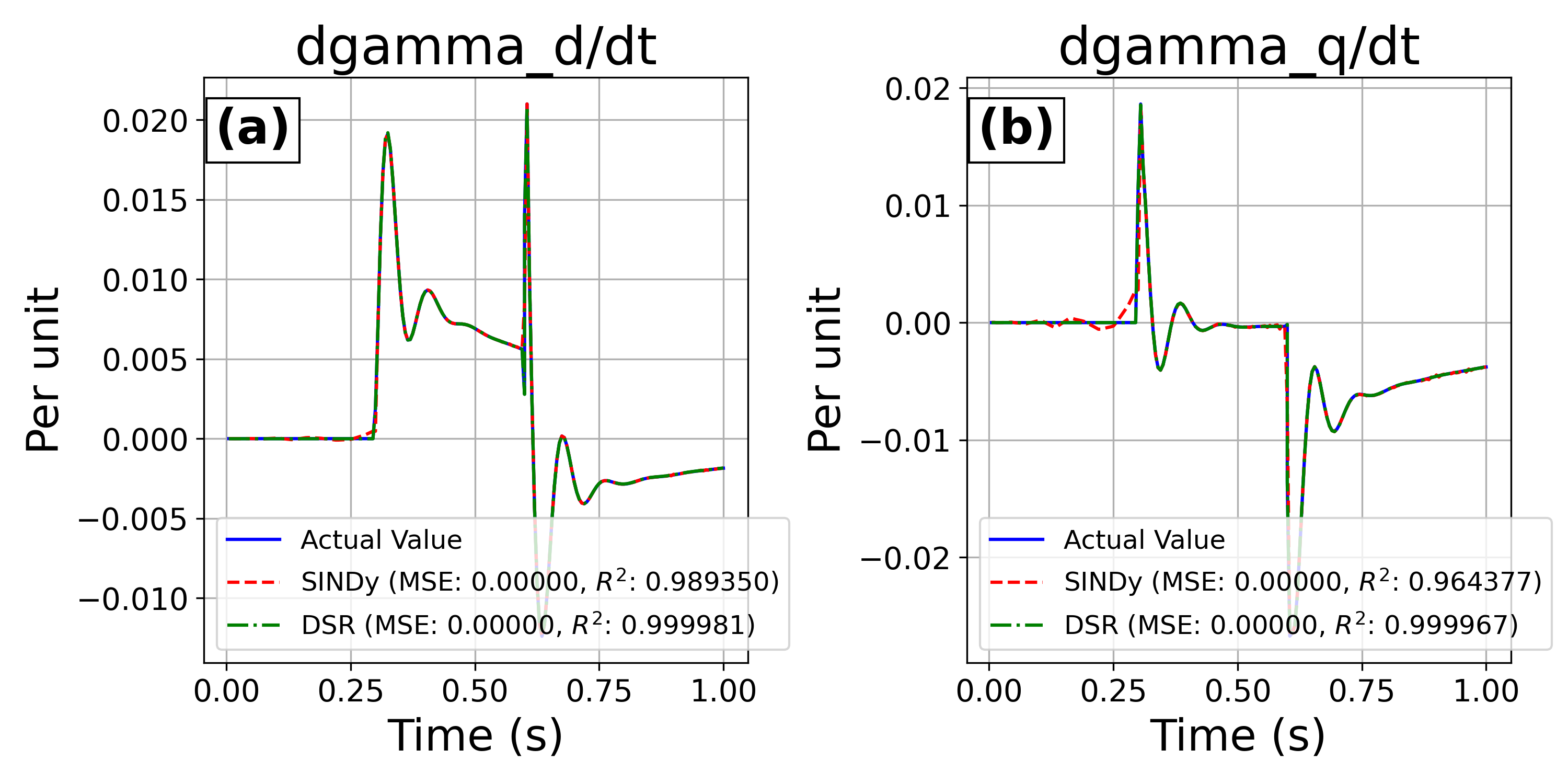}
    \caption{Comparison of the identified dynamics of the inner control in the grid-connected converter under grid-following control mode, using the SINDy and DSR methods. Each subplot illustrates the time derivative of state variables under two distinct disturbances: a change in active power reference (\(p^{ref} = 0.7\)) at \(t=0.3\)s and a change in reactive power reference (\(q^{ref} = 0.2\)) at \(t=0.6\)s. The subplots correspond to each part are as follow; (a)$\frac{d}{dt}{\gamma}_{d}$, and (b)$\frac{d}{dt}{\gamma}_{q}$.}
    \label{inner}
\end{figure}

\subsection{Inner control}
The derived system representation for inner control is highlighted in Figs. \ref{inner}.a and \ref{inner}.b. The SINDy model reaches an \( R^2 \) score of 0.98 in Fig. \ref{inner}.a, reflecting a high level of accuracy. Similarly, DSR illustrates a comparable performance with an \( R^2 \) score of 0.99. Finally, the SINDy model achieves an \( R^2 \) score of 0.96 in Fig. \ref{inner}.b, indicating reasonable performance. However, DSR attains an \( R^2 \) score of 0.99, demonstrating its ability to model this dynamic accurately.

\subsection{Discussion}

The comparative analysis of the SINDy and DSR methodologies presented in this study provides valuable insights into the modeling of grid-following converters in modern power systems. The results indicate that DSR consistently presents better performance compared to SINDy in terms of modeling accuracy, as evidenced by lower mean square errors and higher \(R^2\) scores across various state variables, including the dynamics of the LCL filter, PLL, and outer and inner control loops. This performance can be attributed to DSR's ability to uncover complex, nonlinear relationships within the system dynamics, making it particularly effective for capturing the intricate behaviors of grid-following converters, especially under disturbances such as changes in active and reactive power references.

However, the enhanced accuracy of DSR comes with a significant computational cost. The study reveals that DSR's computational burden is over ten times higher than that of SINDy, which poses a limitation for real-time applications. This trade-off between accuracy and computational efficiency is a critical consideration, particularly in large-scale power systems where real-time monitoring and control are essential. While DSR offers a more precise and interpretable framework for modeling converter dynamics, its high computational requirements may hinder its practical implementation in scenarios requiring rapid model identification and adaptation.

\section{conclusions}
This paper has explored and validated the use of both the SINDy and DSR methodologies for the dynamic model identification of grid-connected converters, with a focus on grid-following control mode. The results indicate that while SINDy is capable of discovering compact models, DSR provides a more accurate and interpretable framework for capturing the nonlinear dynamics of converter-based systems. 
DSR's ability to uncover underlying system behaviors makes it highly effective, particularly in the context of dynamic and complex control systems, such as those found in modern power grids. However, this method may not be practical for real-time model identification due to its computational burden. The findings suggest that DSR could be a crucial tool for optimizing control strategies as renewable energy sources become more prevalent, supporting the transition towards more resilient and flexible grid operations.
Future research should extend these methodologies to larger, more intricate systems and explore their integration with more intricate control mechanisms. Additionally, scaling these approaches to handle larger datasets and more complicated system architectures will be essential in addressing the increasing complexity of grid operations with high renewable penetration.

\bibliographystyle{IEEEtran}
\bibliography{ref}

\end{document}